\documentclass[%
reprint,
superscriptaddress,
 amsmath,amssymb,
 aps,
pra,
longbibliography
]{revtex4-1}

\usepackage{float}
\usepackage{tabularx}
\usepackage{graphicx}% Include figure files
\usepackage{bm}% bold math
\usepackage{tikz}
\usepackage{amsmath}
\usetikzlibrary{arrows.meta}
\usetikzlibrary{decorations.pathreplacing}
\usepackage[percent]{overpic}
\usepackage{hyperref}% add hypertext capabilities
\hypersetup{colorlinks=true, citecolor=blue, urlcolor=blue, linkcolor=blue}
\usepackage{color}
\begin{document}

\title{Low energy paths for octahedral tilting in inorganic halide perovskites}
\author{Johan Klarbring}
 \email{johan.klarbring@liu.se}
 \affiliation{% 
Theoretical Physics Division, \\
Department of Physics, Chemistry and Biology (IFM), 
Link\"{o}ping University, SE-581 83, Link\"{o}ping, Sweden
}%

\date{\today}% It is always \today, today,
             %  but any date may be explicitly specified
\begin{abstract}
Instabilities relating to cooperative octahedral tilting is common in materials with perovskite structures, and in particular in the sub class of halide perovskites. In this work, the energetics of octahedral tilting in the inorganic metal halide perovskites CsPbI$_3$ and CsSnI$_3$ are investigated using first-principles density functional theory calculations. Several low energy paths between symmetry equivalent variants of the stable orthorhombic (\textit{Pnma}) perovskite variant are identified and investigated. The results are in favor of the presence of dynamic disorder in the octahedral tilting phase transitions of inorganic halide perovskites. In particular, one specific type of path, corresponding to an out-of-phase "tilt switch", is found to have significantly lower energy barrier than the others, which indicates the existence of a temperature range where the dynamic fluctuations of the octahedra follow only this type of path. This could produce a time averaged structure corresponding to the intermediate tetragonal (\textit{P4/mbm}) structure observed in experiments. Deficiencies of the commonly employed simple one-dimensional "double well" potentials in describing the dynamics of the octahedra are pointed out and discussed.  
\end{abstract}

\pacs{Valid PACS appear here}% PACS, the Physics and Astronomy
                             % Classification Scheme.
%\keywords{Suggested keywords}%Use showkeys class option if keyword
                              %display desired
\maketitle

\section{Introduction}

The interest in metal-halide materials with perovskite-derived structures has received a tremendous surge in recent years. This is primarily due to the record-level efficiency of photovoltaic devices based on inorganic-organic hybrid perovskites such as methylammonium lead iodide (CH$_3$NH$_3$PbI$_3$ or MAPI) \cite{Kojima2009,Saliba2016}. Among the main factors hindering the use of such devices on a large industrial scale is the rapid degradation due to external influences encountered during operation \cite{Correa-Baena2017}. 

The fully inorganic counterparts to the hybrid halide perovskite, eg.\ CsPbI$_3$ in the case of MAPI, are also increasingly intensively studied. The interest in these systems stem both from being simpler analogs to the hybrid perovskites, but also for their own potential use in optoelectronic devices \cite{Sanehira2017,Cho2017,Jeong2018,Wang2018}. 

\begin{figure}
    \includegraphics[trim={0cm 0cm 0cm 0cm},clip,width=\linewidth]{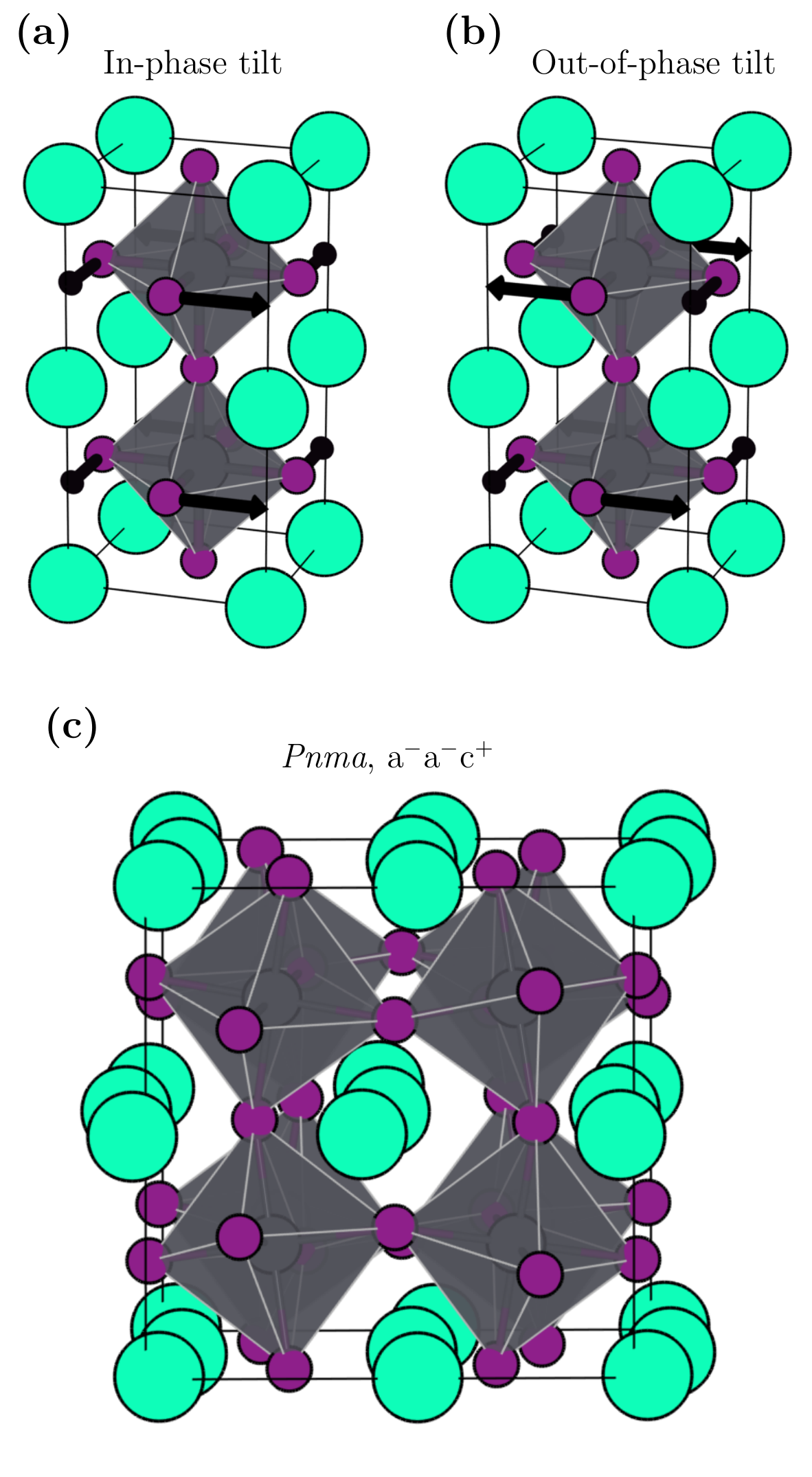}
    \caption{\label{fig:tilts_struct} Illustration of the in-phase (a) and out-of-phase (b) octahedral tilting modes in perovskites and the orthorhombic (\textit{Pnma}) structure with Glazer tilt pattern $a^{-}a^{-}c^{+}$, in the psuedocubic setting (c). Cs and I ions are represented by green and purple spheres, respectively, and the Pb/Sn ions are the black spheres in the center of the octahedra.}
\end{figure}

In general, perovskite, ABX$_3$, based compounds are not found in the aristotype cubic perovskite at low temperature, but instead admit one of many distorted perovskite variants. The most common class of such distortions is the antiferrodistortive (AFD) distortions, which correspond to cooperative tilting of the BX$_6$ octahedral network around one or several of the [100] axes of the cubic perovskite structure. These octahedral tilting instabilities are related to imaginary phonon modes at the M and R points of the 1st Brillouin zone (BZ), corresponding to in-phase and out-of-phase tilts of sequential octahedra around a particular axis, respectively, as illustrated in Figs.\ \ref{fig:tilts_struct} (a) and (b). A host of different combinations of in-phase and out-of-phase tilts are possible \cite{Howard1998}. These are conveniently expressed using Glazer notation \cite{glazer1972} where, as an example, the very common orthorhombic (\textit{Pnma}, Fig.\ \ref{fig:tilts_struct}(c)) structure is denoted a$^{-}$a$^{-}$c$^{+}$, indicating out-of-phase tilts of equal magnitude around the \textit{a} and \textit{b} and an in-phase tilt around the \textit{c} pseudocubic axes.

A representative sequence of AFD transitions in inorganic halide perovskites is the one in CsSnI$_3$ and CsPbI$_3$ \cite{Chung2012,Marronnier2018}. When cooling from the ideal cubic perovskite (\textit{Pm$\bar{3}$m}) $\alpha$-phase, they undergo two octahedral tilting transitions. First, to the tetragonal (\textit{P4/mbm}), a$^{0}$a$^{0}$c$^{+}$, $\beta$-phase and finally to the orthorhombic (\textit{Pnma}) $\gamma$-phase, with tilts around all three pseudocubic axes, a$^{-}$a$^{-}$c$^{+}$. Note that these low temperature AFD perovskite variants are metastable with respect to a non-perovskite orthorhombic $\delta$-phase in both CsSnI$_3$ and CsPbI$_3$ \cite{Chung2012,Marronnier2018}.

Recently, there has been increasing evidence that the high temperature cubic phase of many perovskites is not, in fact, cubic on a local scale \cite{Klarbring2018,Carignano2017,beecher2016,Quarti2016,Bertolotti2017,Patrick2018}. Instead, the cubic phase emerge as a result of static and/or dynamic disorder. The static disorder corresponds to domains with different symmetry broken variants of the cubic perovskite, which can produce a spatially averaged structure of apparent cubic symmetry. The dynamical disorder corresponds to the hopping of the system between different variants of lower symmetry phases. In the case of octahedral tilting transitions, the lower symmetry phases correspond to different tilt configurations of the BX$_6$ octahedral network that are local minima on the potential energy surface (PES). These are symmetrically distributed around the tiltless cubic perovskite. Similar transition mechanisms have recently been proposed also in several non-perovskite materials \cite{Bhattacharya2008,Thomas2013,Carbogno2014,Kadkhodaei2017,Kadkhodaei2018}.

The presence of octahedral tilts in halide perovskites has a large impact on properties related to their optoelectronic performance. For instance, the band gap tends to increase with increased tilting \cite{Wikto2017,Patrick2015,yang2017}. Furthermore, it has been suggested that the long charge carrier lifetime in halide perovskites is related to the presence of dynamic disorder \cite{Munson2018}. Understanding the nature of the octahedral tilting transitions therefore becomes of vital importance.  

It is tempting to rationalize the dynamical disorder of the octahedral tilts in terms of the one dimensional (1D) PES that are found by displacing the atoms in the cubic perovskite structure according to the in-phase or out-of-phase tilting modes. These typically take the form of "double-well" (DW) potentials, and will be referred to as cubic DWs throughout the rest of the paper. However, since the most stable perovskite variant is the $a^{-}a^{-}c^{+}$ structure, the tilt fluctuations likely occur on paths between distinct such $a^{-}a^{-}c^{+}$ variants. Those paths, as opposed to the cubic DWs, do not pass over the high energy cubic structure. They instead pass over lower energy saddle points on the multi-dimensional PES of octahedral tilts \cite{Klarbring2018,Bechtel2018}.  

In this paper I use Density Functional Theory (DFT) to thoroughly investigate the energetics associated with octahedral tilting in inorganic halide perovskites. I identify and investigate several paths with low energy barriers between symmetry equivalent $a^{-}a^{-}c^{+}$ variants. These paths exist as a consequence of the presence of tilts around multiple pseudocubic axes. This implies that the cubic DWs are not the relevant PESs for determining, for example, the rate of fluctuations of the octahedral tilts. I point out that the depth of the cubic DWs are, in fact, inversely related to the energy barrier on certain low energy paths between symmetry equivalent $a^{-}a^{-}c^{+}$ variants.

These low energy paths allow for the system to be dynamically disordered between distinct $a^{-}a^{-}c^{+}$ variants at elevated temperature. One particular path, which corresponds to changing the direction of a single out-of-phase tilt, has a significantly lower energy barrier than the others. In a certain temperature range the octahedral framework will thus fluctuate only along this particular path. This would produce, as a time average, a a$^{0}$a$^{0}$c$^{+}$ structure. This could be related to the experimentally observed intermediate tetragonal a$^0$a$^0$c$^+$ (\textit{P4/mbm}). When the temperature is further increased, several other modes of octahedral fluctuations will become active, and the cubic a$^{0}$a$^{0}$a$^{0}$ phase is the resulting time average structure.

\section{Methods}
\subsection{Computational details}
All DFT calculations were performed in the framework of the projector augmented wave (PAW) \cite{blochl1994} method using the Vienna Ab Initio Simulation Package (VASP) \cite{kresse1996,kresse1996_2,kresse1999}. Exchange and correlation effects were treated using the PBEsol \cite{perdew2008} form of the Generalized Gradient Approximation (GGA). The Kohn-Sham orbitals were expanded in plane waves with a kinetic energy cutoff of 800 eV. The first BZ of the primitive simple cubic perovskite unit cell was sampled on a $8 \times 8 \times 8$ $\Gamma$-centred grid, and was reduced appropriately when larger simulation cells were used. PAW potentials with the Cs(5s5p6s), Pb(6s6p), Sn(4d5s5p), I(5s5p) valence configuration were used. The convergence criterion for the electronic self-consistent field cycles was 10$^{-8}$ eV.

\subsection{Octahedral tilting}
Octahedral tilt patterns in perovskites are conveniently expressed in Glazer notation \cite{glazer1972}, where a tilt pattern is specified by three letters, a, b and c representing the tilts around the three psuedocubic axes. Each letter has an associated superscript '+', '-' or '0',  which indicate an in-phase tilt, an out-of-phase tilt or no tilt, respectively. If the tilt angles around multiple axes are the same, the tilts are represented by the same letter. As an example, the common orthorhombic (\textit{Pnma}) structure is denoted a$^{-}$a$^{-}$c$^{+}$ indicating out-of-phase tilts of equal magnitude around the \textit{a} and \textit{b} and an in-phase tilt around the \textit{c} pseudocubic axes. 

For a given Glazer tilt pattern there exists several distinct but symmetry equivalent variants. These correspond to changing the axis around which a particular tilt is taken, eg.\ changing a$^{-}$a$^{-}$c$^{+}$ to a$^{+}$b$^{-}$b$^{-}$, and/or changing the direction of the tilt from clockwise to counter clockwise or vice versa, eg.\ changing a$^{-}$a$^{-}$c$^{+}$ to a$^{-}$a$^{-}$(-c)$^{+}$. Bechtel and Van der Ven \cite{Bechtel2018} have classified these into rotational and translational variants. The translational variants are those that can be sampled by only changing the octahedral tilts within a fixed strain state, while the rotational variants correspond to different strain states. For a$^{-}$a$^{-}$c$^{+}$, there are 24 distinct variants which can be grouped into 6 rotational equivalents, each with 4 translational variants \cite{Bechtel2018}. The 4 translational variants are related to each other by changing the tilt direction of either the in-phase tilt and/or both the out-of-phase tilts. 

\begin{figure}
    \includegraphics[trim={0cm 0cm 0.49cm 0cm},clip,width=\linewidth]{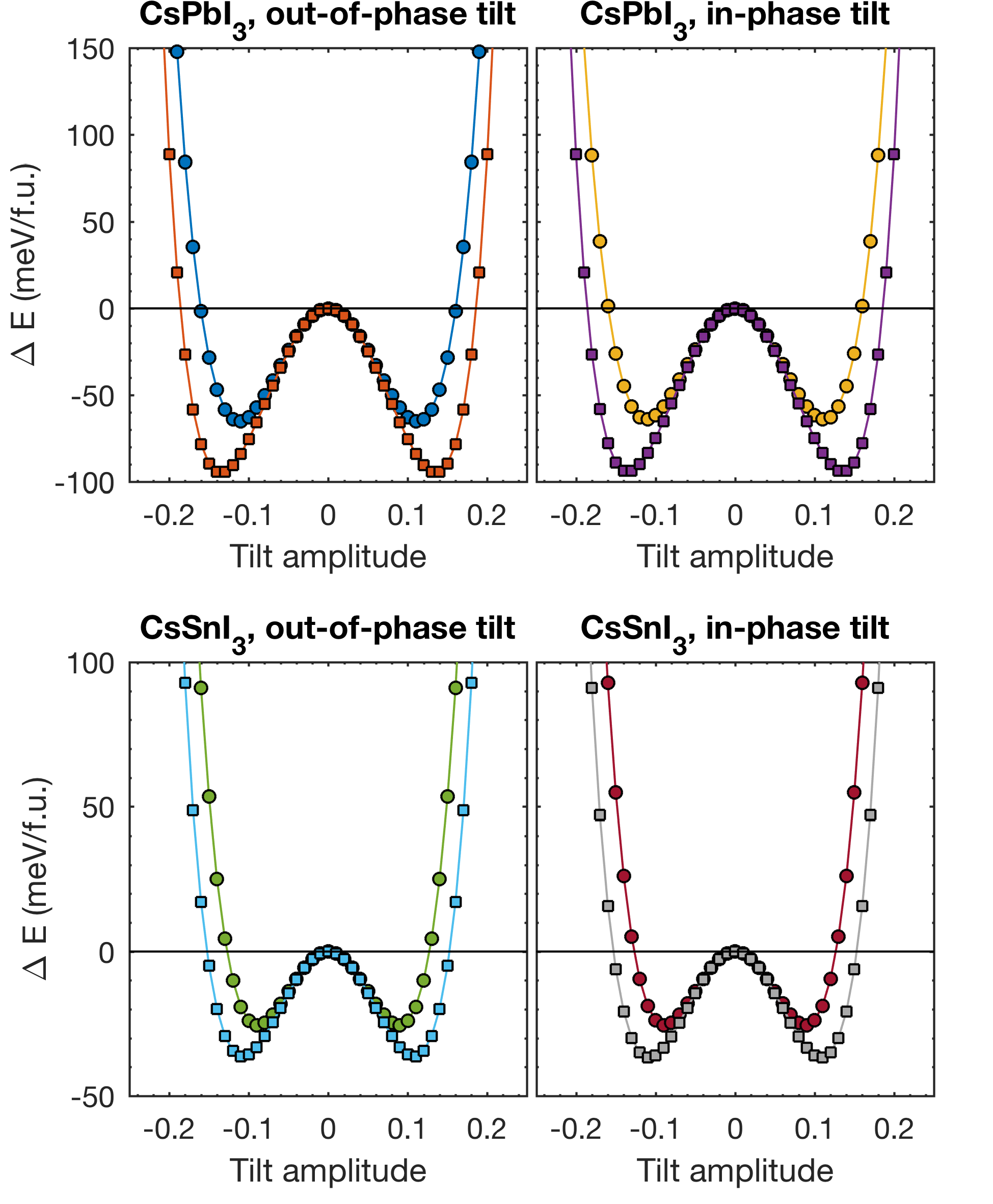}
    \caption{\label{fig:DWs}Potential energy as a function of tilt amplitude for in-phase and out-of-phase tilts in CsPbI$_3$ and CsSnI$_3$. Circles and squares in all four panels represent fixed and tetragonaly relaxed unit cells, respectively.}
\end{figure}
\begin{figure*}
    \includegraphics[trim={0cm 0cm 0cm 0cm},clip,width=\linewidth]{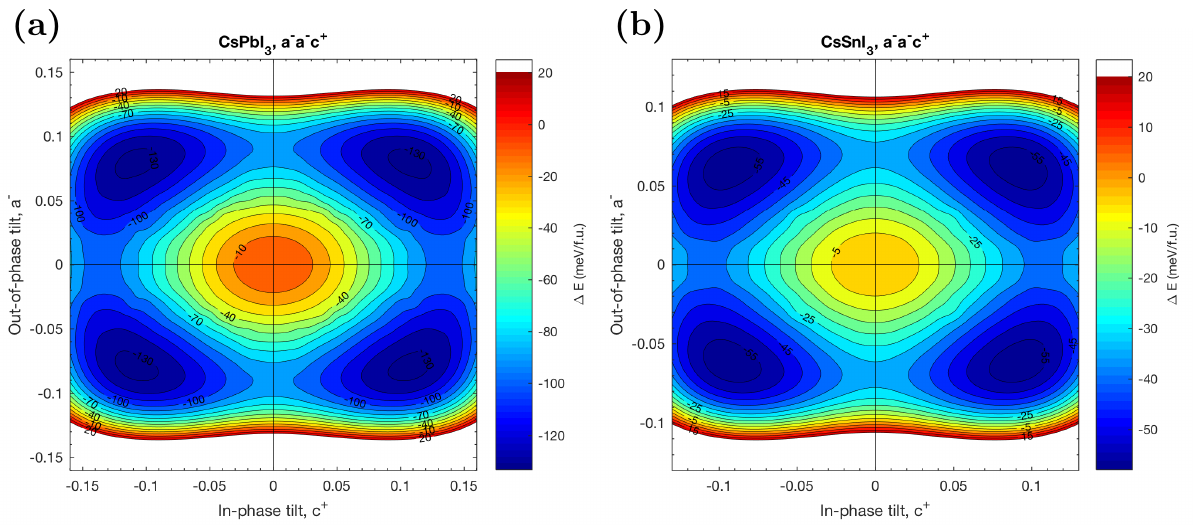}
    \caption{\label{fig:PES} a$^{-}$a$^{-}$c$^{+}$ PES of (a) CsPbI$_3$ and (b) CsSnI$_3$. The x coordinate axis gives the magnitude of the in-phase tilt around the pseudocubic c axis, while the y coordinate axis gives the magnitude of the out-of-phase tilts around the a and b pseudocubic axes. The tilt amplitude is given as the offset of one I ion in units of the lattice constant of the ideal cubic perovskite. Relaxation of Cs ions and lattice parameters are allowed in the symmetry given by the tilt pattern. The energy reference is the cubic phase (origin). The energy contours are spaced by 10 and 5 meV/f.u.\ in (a) and (b), respectively. Note that the energy scales differ in (a) and (b) 
    }
\end{figure*}
PESs as functions of different octahedral tilt combinations were calculated in $2\times2\times2$ supercells (40 atoms). This was done by first displacing the I ions in the ideal cubic structure according to particular combinations of in-phase and out-of-phase tilting modes with varying amplitudes. The positions of the I ions were then frozen, while relaxations of the lattice vectors and the remaining ionic positions where performed. The relaxation was constrained to stay in the symmetry given by the particular octahedral tilting state, which constrains the Pb/Sn ions to stay in the center of the octahedra. The calculations were done on $17\times17$ and $14\times14$ equispaced grids of positive tilt amplitudes for CsPbI$_3$ and CsSnI$_3$, respectively. Relevant symmetries where then exploited by mirroring the energies in the tilt coordinate axes. The results where finally interpolated to a dense grid using cubic polynomials before producing the displayed figures.

Minimum energy paths between symmetry equivalent a$^{-}$a$^{-}$c$^{+}$ tilt configurations were obtained using the climbing image nudged elastic band (CI-NEB) method \cite{Henkelman2000}. Both the standard and the generalized solid state variant (G-SSNEB) \cite{sheppard2012}, which extends the regular NEB method to include the variation of both atomic and unit cell degrees of freedom along the path, were used. No symmetry restrictions on either lattice vectors or ionic positions where imposed along the NEB paths. The initial paths were setup as linear interpolations with seven images between endpoints corresponding to two distinct a$^{-}$a$^{-}$c$^{+}$ variants. The images were then relaxed until the maximum NEB force on all images was less than 5 $\times$ 10$^{-3}$ eV/\AA\ . The distortion of an octahedron along the NEB paths is measured by the parameter \cite{Abakumov2013,Brown1973}:
\begin{equation}
\label{eq:oct_dist}
    \Delta = \frac{1}{6}\sum_{i=1}^6 \left( \frac{d_i-\bar{d}}{\bar{d}} \right)^2,
\end{equation}
where $d_i$ and $\bar{d}$ are the individual and average Pb-I or Sn-I bond length of the octahedron, respectively.   

\section{Results and Discussion}
\subsection{Potential energy surfaces of octahedral tilts}
I start by calculating the cubic DWs of CsPbI$_3$ and CsSnI$_3$, i.e., the 1D PESs  obtained by freezing in in-phase and out-of-phase octahedral tilts in the cubic structure. Fig.\ \ref{fig:DWs} shows the resulting energies for two cases, first simply performing the octahedral tilts while keeping the cubic shape of the supercell and second when allowing a tetragonal relaxation of the lattice. In both cases the results are the expected double well potentials.

It has been suggested \cite{Young2016} that the observed preference of the a$^0$a$^0$c$^+$ structure over a$^0$a$^0$c$^-$ in inorganic halide perovskites can be explained by the higher energy gain associated with in-phase as compared to out-of-phase tilts. This appears to, in fact, be incorrect since the cubic DWs associated with in-phase and out-of-phase octahedral tilts are very similar and the fully relaxed a$^0$a$^0$c$^+$ and a$^0$a$^0$c$^-$ structures have energies within 1 meV/f.u. of each other in both CsPbI$_3$ and CsSnI$_3$. This is in agreement with Ref.\ \cite{Bechtel2018}. Thus, simple energetic considerations are not enough to explain the observed preference of the a$^0$a$^0$c$^+$ structure over a$^0$a$^0$c$^-$ in inorganic halide perovskites. 

Next, I calculate the two dimensional (2D) a$^{-}$a$^{-}$c$^{+}$ PES of CsPbI$_3$ and CsSnI$_3$, i.e., freezing in the same out-of-phase tilt around the a and b axes and an in-phase tilt around the c axis, with varying amplitudes. Relaxation of the Cs ions and lattice vectors were performed within the symmetry determined by the tilt-pattern. The PES of CsPbI$_3$ is shown in Fig.\ \ref{fig:PES} (a). The four minima coincide closely with symmetry equivalent variants of the a$^{-}$a$^{-}$c$^{+}$ structure. The depth of these minima is $\sim 133 $ meV/f.u., which is close to the energy difference of $\sim 137 $ meV/f.u. between the cubic and the \textit{Pnma} structure. The small $\sim 4 $ meV/f.u. discrepancy is due to small relaxations of I ions beyond pure octahedral tilts.
\begin{figure}
    \includegraphics[trim={0cm 0cm 0cm 0cm},clip,width=\linewidth]{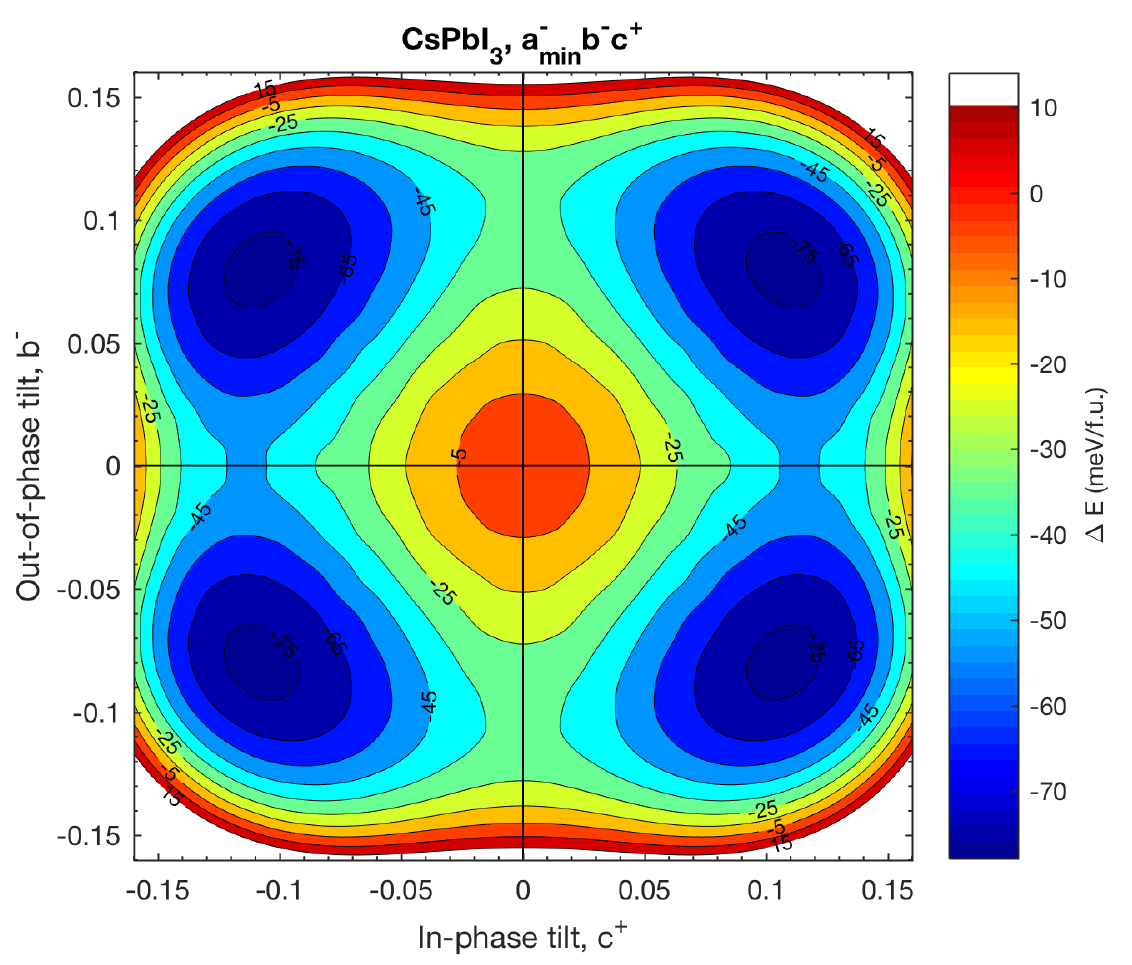}
    \caption{\label{fig:PES_CSI}
    2D octahedral tilting PES in CsPbI$_3$ with the out-of-phase tilt around the a pseudocubic axis fixed to its value in the fully relaxed a$^{-}$a$^{-}$c$^{+}$ structure.  The x coordinate axis gives the magnitude of the in-phase tilt around the pseudocubic c axis, while the y coordinate axis gives the magnitude of the out-of-phase tilt around b pseudocubic axes. Relaxation of Cs ions and lattice parameters are allowed in the symmetry given by the tilt pattern. The energy reference is a a$^-$b$^0$b$^0$ structure, where the magnitude of the a$^-$ tilt is the same as in the fully relaxed a$^-$a$^-$c$^+$ structure. The energy contours are spaced by 10 meV/f.u.\ . 
    }
\end{figure}

Fig.\ \ref{fig:PES} (a) immediately reveals two low energy paths between symmetry equivalent a$^{-}$a$^{-}$c$^{+}$ energy basins. The first, which corresponds to moving horizontally between two basins in Fig.\ \ref{fig:PES} (a), is an in-phase "tilt-switch" in the presence of two out-of-phase tilts, denoted a$^-$a$^-$c$^+$ $\rightarrow$ a$^-$a$^-$(-c)$^+$. The second path, the vertical path between basins in the figure, is the transition a$^-$a$^-$c$^+$ $\rightarrow$ (-a)$^-$(-a)$^-$c$^+$, i.e., switching both the out-of-phase tilts in the presence of an in-phase tilt. The barriers on the paths are $\sim$ 46 and 39 meV/f.u. for the a$^-$a$^-$c$^+$ $\rightarrow$ a$^-$a$^-$(-c)$^+$ and a$^-$a$^-$c$^+$ $\rightarrow$ (-a)$^-$(-a)$^-$c$^+$ transitions, respectively.

The 2D a$^{-}$a$^{-}$c$^{+}$ PES in CsSnI$_3$ is qualitatively similar to CsPbI$_3$, although with shallower energy basins. The energy barriers are 26 and 21 meV/f.u. for the a$^-$a$^-$c$^+$ $\rightarrow$ a$^-$a$^-$(-c)$^+$ and a$^-$a$^-$c$^+$ $\rightarrow$ (-a)$^-$(-a)$^-$c$^+$ transitions, respectively. The lower energy barriers in CsSnI$_3$ compared to CsPbI$_3$ is consistent with the lower phase transition temperatures \cite{Chung2012,Marronnier2018}.

A second 2D PES can be obtained by freezing the out-of-phase tilt around the a axis to its equilibirum value in the a$^-$a$^-$c$^+$ structure, and subsequently varying the remaining out-of-phase and in-phase tilts. The resulting PES for CsPbI$_3$ is shown in Fig.\ \ref{fig:PES_CSI}. Note that the origin in this figure does not correspond to the ideal cubic perovskite with no tilts, but to a a$^-$b$^0$b$^0$-type structure where the magnitude of the a$^-$ tilt is the same as in the a$^-$a$^-$c$^+$ structure. The four minima in Fig.\ \ref{fig:PES_CSI} correspond to different a$^-$a$^-$c$^+$ variants in the same way as in Fig.\ \ref{fig:PES}. Fig.\ \ref{fig:PES_CSI} reveals a path between a$^-$a$^-$c$^+$ basins that is not found in Fig.\ \ref{fig:PES}. This path corresponds to moving vertically between two minima in the figure and is of the form a$^-$a$^-$c$^+$ $\rightarrow$ a$^-$(-a)$^-$c$^+$. The energy barrier on this path obtained from Fig.\ \ref{fig:PES_CSI} is $\sim$ 32 meV/f.u. 

\begin{figure*}
    \includegraphics[trim={0cm 0cm 0cm 0cm},clip,width=\linewidth]{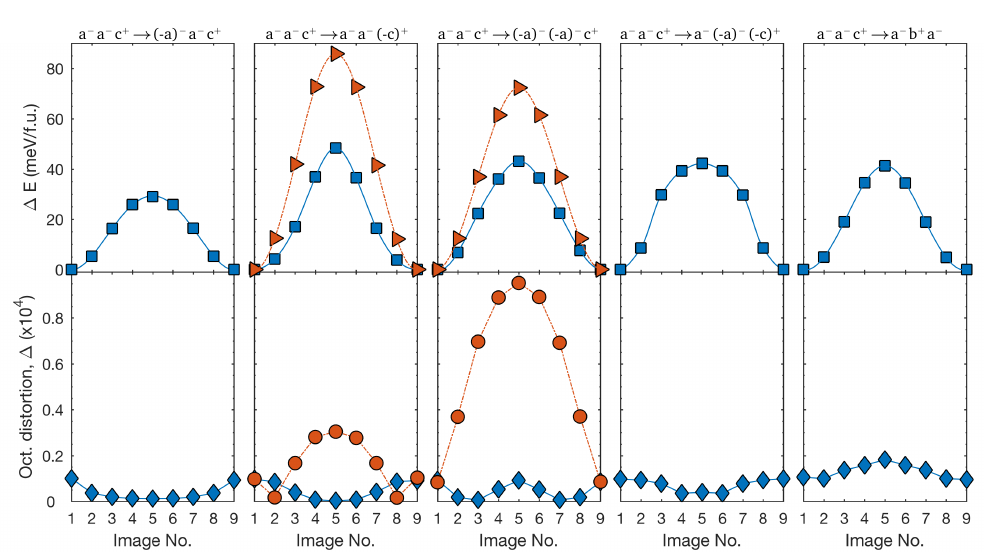}
    \caption{\label{fig:NEBs} (Top row) NEB energy profiles for CsPbI$_3$ on a set of distinct paths between symmetry equivalent variants of the a$^-$a$^-$c$^+$ structure. (Bottom row) Maximum value of the octahedral distortion parameter, $\Delta$ (Eq. \ref{eq:oct_dist}), along the NEB paths. Blue and red markers correspond to a varying and fixed lattice along the NEB path, respectively.
    }
\end{figure*}

\subsection{Minimum energy paths between a$^{-}$a$^{-}$c$^{+}$ variants}
The PESs in Figs.\ \ref{fig:PES} and \ref{fig:PES_CSI} are useful in revealing and visualizing low energy paths between distinct a$^{-}$a$^{-}$c$^{+}$ variants. However, since they are generated from pure octahedral tilts followed by symmetry constrained Cs ion and lattice relaxations, they do not contain all relevant degrees of freedom and do not necessarily reveal the lowest energy paths. To adress this, I have performed G-SSNEB calculations for CsPbI$_3$ and CsSnI$_3$ with endpoints corresponding to distinct fully relaxed a$^{-}$a$^{-}$c$^{+}$ variants. No symmetry constraints were imposed along the paths in order to fully characterize them with respect to both atomic and lattice vector degrees of freedom.

I first consider four paths: the three already mentioned in relation to the PESs of Figs.\ \ref{fig:PES} and \ref{fig:PES_CSI}, i.e. a$^-$a$^-$c$^+$ $\rightarrow$ a$^-$a$^-$(-c)$^+$, a$^-$a$^-$c$^+$ $\rightarrow$ (-a)$^-$(-a)$^-$c$^+$ and a$^-$a$^-$c$^+$ $\rightarrow$ a$^-$(-a)$^-$c$^+$ and a fourth path of the form a$^-$a$^-$c$^+$ $\rightarrow$ a$^-$(-a)$^-$(-c)$^+$, i.e. simultaneously switching one in-phase and one out-of-phase tilt. 

The G-SSNEB energy profiles on these paths for CsPbI$_3$ are shown in Fig.\ \ref{fig:NEBs} and the energy barriers for both CsPbI$_3$ and CsSnI$_3$ are listed in Table \ref{table:energies}. Table \ref{table:energies} also lists the energies, with a$^-$a$^-$c$^+$ as reference, for a chosen set of simple Glazer tilt patterns. One can see that the energy barriers on each of the four paths correspond to the energy of one of the simple tilt patterns. For example, the energy barrier of the path a$^-$a$^-$c$^+$ $\rightarrow$ a$^-$(-a)$^-$c$^+$ corresponds to the energy difference between a$^-$a$^-$c$^+$ and 
a$^-$b$^0$c$^+$ (or equivalently a$^-$b$^+$c$^0$).

Upon further investigations of the atomic positions and strain states of the saddle point structures on each of the four paths, it was found that they indeed correspond to one of the simple tilt configurations in Table \ref{table:energies}, where the saddle point structure for each path is listed.

The two paths a$^-$a$^-$c$^+$ $\rightarrow$ a$^-$a$^-$(-c)$^+$ and a$^-$a$^-$c$^+$ $\rightarrow$ (-a)$^-$(-a)$^-$c$^+$ have endpoints that belong to the same strain state, i.e. they are translational variants of each other \cite{Bechtel2018}. This makes it possible to calculate NEB paths in the fixed lattice shape corresponding to the endpoints, i.e. where only the atomic positions are active degrees of freedom. These NEB energy profiles are displayed in Fig.\ \ref{fig:NEBs}. The energy barriers on these paths are seen to drastically increase, indicating that the rearrangement of the lattice vectors is crucial in order to get low energy paths. This can partially be understood by looking at the distortion of the octahedra along the NEB paths. The bottom row of Fig.\ \ref{fig:NEBs} shows the maximum value of the octahedral distortion parameter, $\Delta$ (Eq.\ \ref{eq:oct_dist}), along the NEB paths. It is seen that when the lattice vectors are allowed to change, the octahedra tend to stay largely undistorted, while for a fixed lattice, the octahedra will be forced distort which contributes to a significantly higher energy at the saddle point.

A distinctively different path type is of the form a$^-$a$^-$c$^+$ $\rightarrow$ a$^-$b$^+$a$^-$, i.e., an in-phase tilt axis is transformed into out-of-phase by switching every second octahedron, and similarly changing one of the out-of-phase tilt axes to an in-phase tilt. Expect for the lowest energy a$^-$a$^-$c$^+$ $\rightarrow$ a$^-$(-a)$^-$c$^+$ path, the energy barrier on this path, is similar to the other investigated paths in both CsSnI$_3$ and CsPbI$_3$. The saddle point structure on this path does not correspond to one of the simple Glazer tilt patterns but is more complicated. It has the out-of-phase tilt around the a axis still present and a more complicated in-phase-type tilts, with varying amplitudes for the two layers of octahedra, around both the b and c axes. 

\begin{table*}
%\centering
    \begin{tabular}{p{0.15\linewidth}p{0.11\linewidth}p{0.11\linewidth}p{0.01\linewidth}p{0.22\linewidth}p{0.10\linewidth}p{0.10\linewidth}p{0.15\linewidth}}
        \hline \hline
        \multicolumn{3}{c}{Structural energies} & &\multicolumn{3}{c}{Octahedral tilting energy barriers}  \\
        \cline{1-3}\cline{5-8} 
        \textbf{Structure}  & \multicolumn{2}{c}{\textbf{Energy (meV/f.u.)}} & & \textbf{Path type}  & \multicolumn{2}{c}{\textbf{E$_{barrier}$ (meV/f.u.)}} & \textbf{Saddle point} \\
        & \hspace{8 mm} CsPbI$_3$ & \hspace{2 mm} CsSnI$_3$ & & & CsPbI$_3$ & CsSnI$_3$ \\
         \cline{1-3}\cline{5-8}                    
       a$^0$a$^0$a$^0$ (\textit{Pm$\bar{3}$m}) & \hspace{4 mm} 137    &\hspace{5 mm} 60    &  &a$^-$a$^-$c$^+$ $\rightarrow$ a$^-$a$^-$(-c)$^+$     &\hspace{1 mm} 48 & 28 & a$^-$a$^-$c$^0$ \\
       a$^0$a$^0$c$^+$ (\textit{P4/mbm})       & \hspace{5 mm} 43  &\hspace{5 mm} 23  &  &a$^-$a$^-$c$^+$ $\rightarrow$ (-a)$^-$(-a)$^-$c$^+$  &\hspace{1 mm} 43 & 23 & a$^0$a$^0$c$^+$ \\
       a$^0$a$^0$c$^-$ (\textit{I$4/$mcm})     & \hspace{5 mm} 42  &\hspace{5 mm} 24  &  &a$^-$a$^-$c$^+$ $\rightarrow$ (-a)$^-$a$^-$c$^+$     &\hspace{1 mm} 29 & 15 & a$^0$b$^-$c$^+$ (a$^-$b$^+$c$^0$)\\
       a$^-$a$^-$c$^0$ (\textit{Imma})         & \hspace{5 mm} 48  &\hspace{5 mm} 28  &  &a$^-$a$^-$c$^+$ $\rightarrow$ a$^-$(-a)$^-$(-c)$^+$        &\hspace{1 mm} 42 & 24 & a$^-$b$^0$b$^0$ (a$^0$a$^0$c$^-$)\\
       a$^-$b$^+$c$^0$ (\textit{Cmcm})         & \hspace{5 mm} 29 &\hspace{5 mm} 15  &  &a$^-$a$^-$c$^+$ $\rightarrow$ a$^-$b$^+$a$^-$        &\hspace{1 mm} 41 & 23 \\
       a$^-$a$^-$c$^+$ (\textit{Pnma})         & \hspace{5 mm} 0 &\hspace{5 mm} 0  &  &    &  \\
        \hline \hline
\end{tabular}
\caption{\label{table:energies} Left: Structural energies, with a$^-$a$^-$c$^+$ (\textit{Pnma}) as reference, for a set of tilt configurations in CsPbI$_3$ and CsSnI$_3$. Right: Energy barriers for different types of octahedral tilting paths (see text for details) between symmetry equivalent variants of the a$^-$a$^-$c$^+$ structure in CsPbI$_3$ and CsSnI$_3$.}
\end{table*}
\subsection{Discussion}
It is clear from the previous sections that there exist several distinct types of low energy paths between symmetry equivalent a$^{-}$a$^{-}$c$^{+}$ variants in both CsPbI$_3$ and CsSnI$_3$. This suggests that, as the temperature increases, the system will eventually become dynamically disordered among these $a^{-}a^{-}c^{+}$ variants. Such dynamic disorder would produce phases of higher symmetry in the long time average. This is in line with recent proposals of dynamic disorder in the octahedral tilting transitions of halide perovksites \cite{Carignano2017,beecher2016,Bertolotti2017}. The Cs ions are displaced in different patterns from their positions in the ideal cubic structure depending on the particular $a^{-}a^{-}c^{+}$ variant. This implies that a dynamic disorder of the octahedra along the type of paths described above would imply that the Cs ions also becomes dynamically disordered. 

The a$^-$a$^-$c$^+$ $\rightarrow$ a$^-$(-a)$^-$c$^+$ (or equivalently a$^-$a$^-$c$^+$ $\rightarrow$ (-a)$^-$a$^-$c$^+$) path has significantly lower energy barrier than the others, which implies the existence of a temperature range where fluctuations of the octahedra only along this type of path are active. Since the in-phase tilt is still present during such fluctuations, while the out-of-phase tilts will average out to zero, a a$^0$a$^0$c$^+$ (\textit{P4/mbm}) structure would emerge as a time average. This could be related to the existence of an intermediate tetragonal (\textit{P4/mbm}) structure in several inorganic halide perovskites \cite{Chung2012,Fabini2016,Marronnier2018}, in spite of the fact that a$^0$a$^0$c$^+$ is not energetically preferable compared to other tilt configurations (see Table.\ \ref{table:energies}). 

From PESs similar to Fig.\ \ref{fig:PES}, but in a fixed a$^-$a$^-$c$^+$ strain state, Bechtel and van der Ven \cite{Bechtel2018} pointed out that, out of the two paths with endpoints in a fixed a$^-$a$^-$c$^+$ strain state, a$^-$a$^-$c$^+$ $\rightarrow$ (-a)$^-$(-a)$^-$c$^+$ has a lower barrier than a$^-$a$^-$c$^+$ $\rightarrow$ a$^-$a$^-$(-c)$^+$. If the system is restricted to a fixed shape such that only these two paths are available, a dynamic disorder of the octahedral tilts would, in the same way as described above, produce an a$^0$a$^0$c$^+$ (\textit{P4/mbm}) structure in a restricted temperature range.

The energy barriers listed in Table \ref{table:energies} are significantly lower than the depth of the 1D cubic DW potentials in Fig.\ \ref{fig:DWs}. This is because the low energy paths pass over saddle points on a higher dimensional PES corresponding to tilted structures, while the 1D cubic DWs pass over the high energy tiltless cubic structure. In fact, there is an inverse relationship between the depth of the cubic 1D double wells and the energy barrier on certain low energy paths between $a^{-}a^{-}c^{+}$ variants. As an example, the saddle point on the a$^-$a$^-$c$^+$ $\rightarrow$ (-a)$^-$(-a)$^-$c$^+$ path, is the a$^0$a$^0$c$^+$ structure, i.e., the bottom of the (tetragonally relaxed) in-phase 1D double well in Fig.\ \ref{fig:DWs}. Thus if, for a fixed energy difference between the cubic and the a$^-$a$^-$c$^+$ structure, the 1D double well gets deeper this corresponds to a lower barrier of the a$^-$a$^-$c$^+$ $\rightarrow$ (-a)$^-$(-a)$^-$c$^+$ path. The depth of the 1D cubic DW potentials can therefore not, on their own, be used as estimates of the relevant energy barriers for dynamical octahedral tilting.   

In reality, the picture of a perfect defect free bulk that is dynamically fluctuating along particular octahedral tilting paths is very likely to simplistic. Recent studies indicate the prevalence of ferroelastic twin boundaries \cite{Hermes2016,Strelcove2017,Bertolotti2017,Patrick2018,Liu2018} in halide perovskites. Bertlotti et al. \cite{Bertolotti2017} demonstrated that the apparent tetragonal and cubic phases, as probed using x-ray diffraction, can result from domains of twinned orthorhombic a$^-$a$^-$c$^+$ variants. They further propose that dynamic cooperative rotations of the octahedral framework are present. The details of how the calculated energy barriers in this work relates to the presence of such twin boundaries is beyond the scope of this paper. It is reasonable to think, however, that the existence of the low energy paths of the type calculated in this work should be beneficial for the formation and rearrangement of such twin boundaries. 

\section{Summary and Conclusion}
To summarize, I have shown that there are several low energy paths between symmetry equivalent a$^-$a$^-$c$^+$ octahedral tilt configurations in inorganic metal halide perovskites. These paths avoid the high energy titless cubic perovskite structure. They instead pass over low energy saddle points on the PES where octahedral tilts are still present in the structure. The energy barriers on the paths are significantly lower than those derived from the commonly employed 1D "double-well" (DW) potentials obtained by freezing in the in-phase or out-of-phase tilts directly in the cubic structure ("cubic DWs"). 

The existence of the low energy paths are in favor of the presence of dynamic disorder of the octahedral tilts at elevated temperature in halide perovskites. Such dynamic disorder would produce time avergaed structures of higher symmetry. In both CsPbI$_3$ and CsSnI$_3$ one particular type of path has significantly lower energy barrier than the others. This path corresponds to switching the out-of-phase tilt around one axis, i.e. a path of the type a$^-$a$^-$c$^+$ $\rightarrow$ (-a)$^-$a$^-$c$^+$ ( or equivalently a$^-$a$^-$c$^+$ $\rightarrow$ a$^-$(-a)$^-$c$^+$). This suggests that the octahedra would become disordered only along this particular type of path in a limited temperature range. In the long time average, the out-of-phase tilts in the system will then average out to zero, yielding an average a$^0$a$^0$c$^+$ structure. This could be related to experimentally observed tetragonal a$^0$a$^0$c$^+$ (\textit{P4/mbm}) structure as an intermediate phase. As temperature is increased further, tilt fluctuations along other types of paths will eventually become active and a titless cubic phase will emerge as the time average. 

\section{Acknowledgements}
The author thanks Prof.\ S.\ I.\ Simak for useful discussions and comments on an early version of the manuscript. Support from the Swedish Research Council (VR) (Project No. 2014-4750) and the Centre in Nano Science and Nano Technology (CeNano) at Link{\"o}ping University is acknowledged. The computations were performed on resources provided by the Swedish National Infrastructure for Computing (SNIC) at the PDC Centre for High Performance Computing (PDC-HPC) and the National Supercomputer Center (NSC).

\bibliographystyle{apsrev4-1}
%\bibliography{refs.bib}  
%merlin.mbs apsrev4-1.bst 2010-07-25 4.21a (PWD, AO, DPC) hacked
%Control: key (0)
%Control: author (72) initials jnrlst
%Control: editor formatted (1) identically to author
%Control: production of article title (-1) disabled
%Control: page (0) single
%Control: year (1) truncated
%Control: production of eprint (0) enabled
%

\end{document}